\documentclass[aps,prl,twocolumn,superscriptaddress,showpacs]{revtex4}
\usepackage{graphicx,amsmath,amssymb,bm,epsfig}

\newcommand{\be}{\begin{equation}}
\newcommand{\ee}{\end{equation}}
\newcommand{\bea}{\begin{eqnarray}}
\newcommand{\eea}{\end{eqnarray}}
\newcommand{\fgies}[3]{\mbox{\raisebox{#3} 
{\epsfig{file=#1,scale=#2,clip=true}}~}}
\newcommand{\pf}{p_{\rm F}}
\newcommand{\ef}{\varepsilon_{\rm F}}
\newcommand{\ntot}{N_{\rm tot}}

\begin{document}

\title{Finite-size and confinement effects in spin-polarized trapped 
Fermi gases}

\author{Mark Ku}
%\email[E-mail:~]{mkmark@triumf.ca}
\affiliation{Department of Physics and Astronomy, University of British 
Columbia, Vancouver, BC V6T 1Z1, Canada}
\affiliation{TRIUMF, 4004 Wesbrook Mall, Vancouver, BC, V6T 2A3, Canada}
\author{Jens Braun}
%\email[E-mail:~]{braun@triumf.ca}
\affiliation{TRIUMF, 4004 Wesbrook Mall, Vancouver, BC, V6T 2A3, Canada}
\author{Achim Schwenk}
%\email[E-mail:~]{schwenk@triumf.ca}
\affiliation{TRIUMF, 4004 Wesbrook Mall, Vancouver, BC, V6T 2A3, Canada}

%\date{\today}

\begin{abstract}
We calculate the energy of a single fermion interacting resonantly
with a Fermi sea of different-species fermions in anisotropic traps,
and show that finite particle numbers and the trap geometry impact
the phase structure and the critical polarization. Our 
findings contribute to understanding some experimental discrepancies
in spin-polarized Fermi gases as finite-size and confinement effects.
\end{abstract}

\pacs{03.75.Ss, 05.30.Fk, 03.75.Hh, 71.10.Ca}

\maketitle

Experiments with spin-polarized Fermi 
gases~\cite{Zwier1,Hulet1,Zwier2,Shin1,Hulet2,Schunck,Shin2}
enable a unique exploration of superfluidity and universal 
properties in strongly-interacting asymmetric Fermi systems.
There are very exciting experimental results of the 
MIT~\cite{Zwier1,Zwier2,Shin1} and Rice 
University~\cite{Hulet1,Hulet2} groups, however
with differences in the observed phase structure and
the critical polariztion. In this Letter, we provide a
first microscopic explanation of the MIT-Rice differences:
The particle number and the trap geometry affect the
interaction energy in the normal polarized phase and
this impacts the limit of superfluidity in traps.

The MIT experiment~\cite{Zwier1,Zwier2,Shin1} observed phase
separation in the trap, with equal densities in the core,
surrounded by a partially-polarized shell and an outer region
of normal majority fermions. The study of vortices~\cite{Zwier1},
in-situ density distributions~\cite{Zwier2}, and the
condensate fraction~\cite{Zwier1,Shin1} established
a critical polarization $P_c = (N_{\uparrow}-N_{\downarrow})/\ntot
= 0.70(3)$ for the superfluid 
phase to exist. These results were obtained in a
harmonic trap with cylindrical symmetry ($\omega_x=\omega_y=
\alpha \omega$; $\omega_z=\omega$), with aspect ratio $\alpha
\sim 5$, and total particle numbers $\ntot = N_{\uparrow}+
N_{\downarrow} \sim 10^6-10^7$. The Rice experiment~\cite{Hulet1,Hulet2}
also observed phase separation, with a fully-paired core 
surrounded by normal majority fermions, but with a sharp phase
boundary and extremely thin partially-polarized shell for low 
temperatures, and the unpolarized core exists to high 
polarization $P_c \gtrsim 0.9$. These experiments are in a 
highly elongated trap, with aspect ratio $\alpha \sim 35-45$,
and for lower $\ntot \sim 10^5$. For these conditions, the
core deformation and the double-peak structure in the
axial density imply a breakdown of the local-density 
approximation (LDA) in Refs.~\cite{Hulet1,Hulet2}.
%For these conditions, the
%deformation of the superfluid core (different from the trap 
%shape) and the double-peak structure in the axial 
%density imply a breakdown of the local-density 
%approximation (LDA) in Refs.~\cite{Hulet1,Hulet2}.

The critical polarization is influenced by the energy of the
competing normal polarized phase.
For large asymmetries, this is governed by the energy of a 
spin-down fermion interacting resonantly with a spin-up Fermi
sea, which is a universal function of $\alpha$ and $N$. We 
study the energy of this spin-down fermion, the so-called
Fermi polaron, in anisotropic traps for different
particle numbers and show that the experimental differences
can be understood partially based on our microscopic results.
This strongly-interacting Fermi polaron provides
insights to problems in condensed-matter systems, with lower
dimensions playing the role of trap-geometry effects, as well
as to nuclear physics, where neutron-rich nuclei exhibit
neutron skins~\cite{NSAC}, with neutron/proton densities
similar to the spin densities in resonantly-interacting cold atoms.

{\it Uniform system.} The polaron energy $E$ 
was calculated variationally for the uniform system including 
one-particle--one-hole excitations (1p1h)~\cite{Chevy1,Chevy2}
and estimated in Ref.~\cite{BF}. 
This leads to a Schwinger-Dyson equation,
$G^{-1}(E)=0$, or diagrammatically:
\begin{multline}
E=G_0^{-1}(E,{\bf p} = {\bf 0})
=(\,\fgies{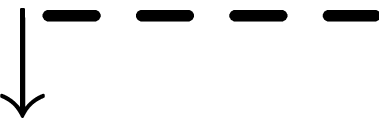}{0.2}{-1.1mm})^{-1} \\[1.5mm]
= \, \fgies{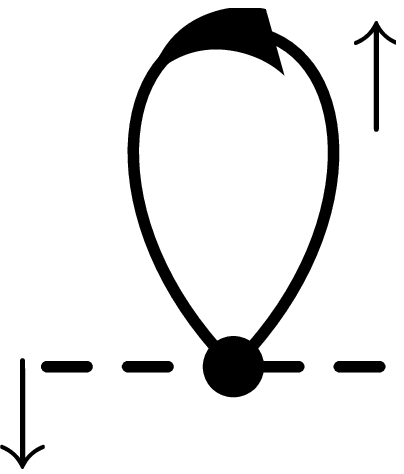}{0.19}{-1.1mm} \, + \: 
\fgies{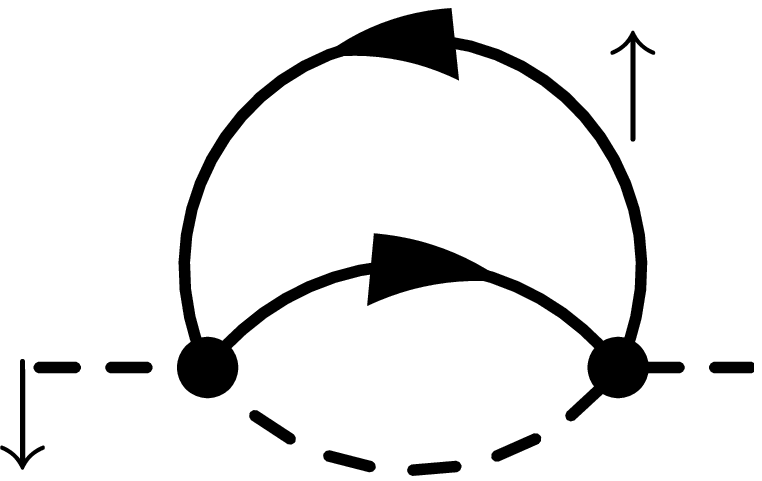}{0.19}{-1.4mm} \, + \: 
\fgies{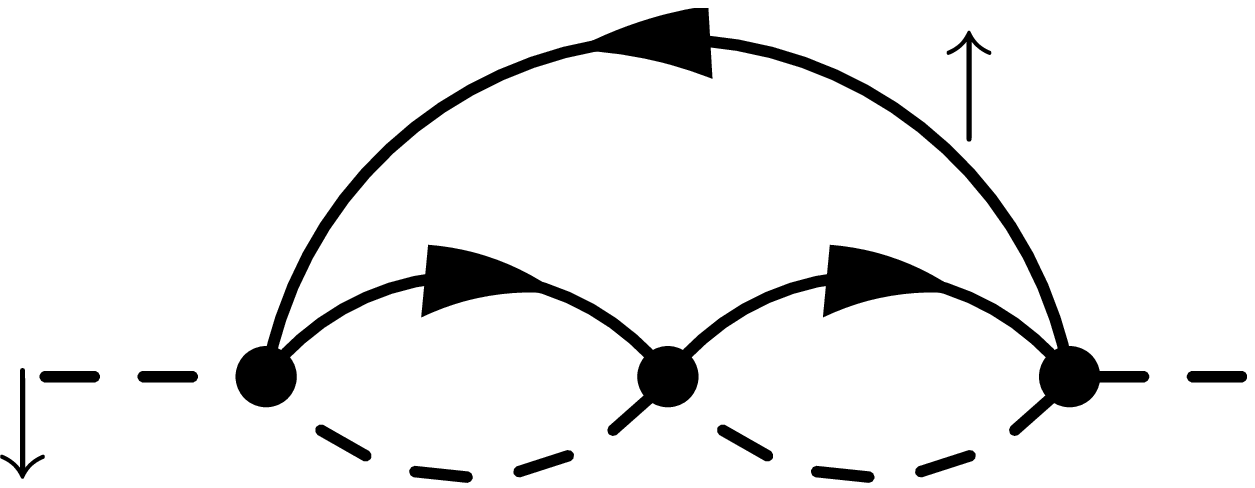}{0.19}{-1.4mm} \, + \: \cdots \,.
\label{diags}
\end{multline}
where $G$ ($G_0$) is the full (noninteracting) spin-down 
propagator with momentum ${\bf p} = {\bf 0}$. For large 
S-wave scattering lengths, $1/a_{\rm s} =0$, the energy
is universal, $E = \mu_\downarrow = \eta \, \pf^2/(2m)$, with
Fermi momentum $\pf$. The self-consistent solution
to Eq.~(\ref{diags}) yields $\eta = -0.607$~\cite{Chevy1}.

This energy gain constrains the equation of state for large
asymmetries, and thus the existence of partially-polarized
phases and the critical polarization: The variational $\eta$
is lower than the maximal stress (from $\mu_\uparrow - \mu_\downarrow
\leqslant 2 \Delta$) for stability of the superfluid phase, 
$\mu_\downarrow/\mu_\uparrow \geqslant -0.09(3)$, and this
requires the existence of at least one nontrivial 
partially-polarized phase in the uniform system~\cite{Chevy1,BF}.
In LDA with Eq.~(\ref{energy}), $\eta=-0.607$ leads to $P_c = 0.74$
and a critical density ratio $x_c = n_{\downarrow}/n_{\uparrow}
= 0.47$~\cite{Lobo}, which are in good agreement with $P_c$
of the MIT experiment~\cite{Zwier1,Zwier2,Shin1} and with
a tomography measurement of $x_c \approx 0.47$~\cite{Shin2}.
Finally, the variational 1p1h $\eta$ value agrees
very well with Monte-Carlo (MC) results~\cite{Lobo,CR,PS,Pilati},
and 2p2h contributions were shown to be small~\cite{Combescot2}.

{\it Basic formalism.} The strongly-interacting Fermi gas in a
harmonic-oscillator trap is given by the Hamiltonian
\begin{multline}
H = \sum_{{\bf n},\sigma} \varepsilon_{\bf n} \, a^{\dag}_{{\bf n},\sigma} \,
a_{{\bf n},\sigma} + \\
\sum_{{\bf n}_{\uparrow},{\bf n}_{\downarrow},
{\bf n}'_{\uparrow},{\bf n}'_{\downarrow}}
\langle {\bf n}'_{\uparrow} , {\bf n}'_{\downarrow} | V | {\bf n}_{\uparrow} ,
{\bf n}_{\downarrow} \rangle \:
a^{\dag}_{{\bf n}'_{\uparrow},\uparrow} \, a^{\dag}_{{\bf n}'_{\downarrow},
\downarrow} \, a_{{\bf n}_{\downarrow},\downarrow} \, a_{{\bf n}_{\uparrow},
\uparrow} \,,
\end{multline}
where $\varepsilon_{{\bf n}} = \alpha \omega(n_{x}+n_{y}+1)+\omega (n_{z}+1/2)$
are harmonic oscillator energies ($\hbar = 1$).
The operator $a_{{\bf n}, \sigma}$ annihilates a particle with
spin $\sigma=\uparrow,\downarrow$ in a state with quantum numbers 
${\bf n}=(n_{x},n_{y},n_{z})$. We use a contact interaction regulated by
separable cutoff functions in momentum space,
\be
\langle {\bf p} | V | {\bf p}' \rangle = C(\Lambda) \, e^{-(p^2+p'^2)/
\Lambda^2} \text{ with } C(\Lambda) = \frac{4\pi/m}{
\frac{1}{a_{\rm s}} - \frac{\Lambda}{\sqrt{2\pi}}} \,,
\ee
where ${\bf p}$, ${\bf p}'$ are incoming/outgoing relative 
momenta, $m$ is the fermion mass and $\Lambda$ a momentum cutoff.
In this case, the harmonic-oscillator matrix elements can be 
expressed as a sum over separable functions
$F({\bf n}_{1},{\bf n}_{2},{\bf S})$,
\be
\langle {\bf n}_{1} , {\bf n}_{2} | V | {\bf n}_{3} , {\bf n}_{4} \rangle
= C(\Lambda) \sum_{{\bf S}} F({\bf n}_{1},{\bf n}_{2},{\bf S})
F({\bf n}_{3},{\bf n}_{4},{\bf S}) \,,
\ee
with center-of-mass quantum numbers ${\bf S}$,
$F({\bf n}_{1},{\bf n}_{2},{\bf S}) = \prod_{i=x,y,z} (m\omega_{i})^{1/4} \,
\widetilde{F}\bigl(n_{1_{i}},n_{2_{i}},S_{i},\lambda_i = 
\frac{\sqrt{m \omega_i/2}}{\Lambda}\bigr)$,
and the dimensionless function $\widetilde{F}$ is given by
\begin{align}
&\widetilde{F}(n_{1_{i}},n_{2_{i}},S_{i},\lambda_i)
= (-1)^{n_{2_{i}}} \, \biggl( \frac{1}{2\pi} \biggr)^{1/4}
\sqrt{\frac{n_{1_{i}}! \, n_{2_{i}}!}{2^{n_{1_{i}}+n_{2_{i}}} \, S_{i}!}}
\nonumber \\
&\times \frac{(n_i-1)!!}{n_i!} 
\frac{(1-2\lambda^{2}_{i})^{n_i/2}}{
(1+2\lambda_{i}^{2})^{(n_i+1)/2}} \: f(n_i,S_{i},n_{2_{i}}) \,,
\end{align}
where the relative quantum numbers $n_i = n_{1_{i}}+n_{2_{i}}-S_i$
have to be even and positive, and one has for $n_{2_{i}} \leqslant n_i$ 
\be
f(n_{i},S_{i},n_{2_{i}})=\biggl(
\begin{array}{c}
n_{i} \\
n_{2_{i}}
\end{array}
\biggr)
\: _{2}F_{1}(-n_{2_{i}},-S_{i},1-n_{2_{i}}+n_{i},-1) \,,
\ee
with hypergeometric function $_{2}F_{1}$,
and for $n_{2_{i}} > n_i$
\begin{multline}
f(n_{i},S_{i},n_{2_{i}})=(-1)^{n_{2_{i}}+n_{i}} \,
\biggl(
\begin{array}{c}
S_{i} \\
n_{2_{i}} - n_{i}
\end{array}
\biggr) \\
\times \: _{2}F_{1}(-n_{i},n_{2_{i}}-n_{i}-S_{i},1+n_{2_{i}}-n_{i},-1) \,.
\end{multline}

{\it Polaron energy.}
Following the variational Ansatz of Refs.~\cite{Chevy1,Chevy2},
we calculate the energy $E$ of the spin-down fermion,
including 1p1h excitations in the wave function,
\be
|\psi\rangle = \phi_{0} \, |\Omega\rangle + \sum_{{\bf m},{\bf h},{\bf p}}
\phi_{{\bf m},{\bf h},{\bf p}} \, |{\bf m},{\bf h},{\bf p}\rangle \,,
\label{Ansatz}
\ee
where $|\Omega\rangle$ denotes the Fermi sea with 
the spin-down particle in the ${\bf n}={\bf 0}$ level~\footnote{The
restriction to ${\bf n}={\bf 0}$ (justified for large $N$)
enables our still involved numerical solution.},
%enables our still involved numerical solution (12 nested sums).},
and $|{\bf m},
{\bf h},{\bf p}\rangle$ consist of a spin-up fermion in ${\bf h}$ 
excited to a level ${\bf p}$ above the Fermi energy $\ef$, and the
spin-down particle occupies the 
level ${\bf m}$. Therefore, the sum over ${\bf h}$ is restricted
to occupied states, whereas ${\bf p}$ is over unoccupied states 
above $\ef$.

\begin{figure}[t]
\begin{center}
\includegraphics[scale=0.45,clip=]{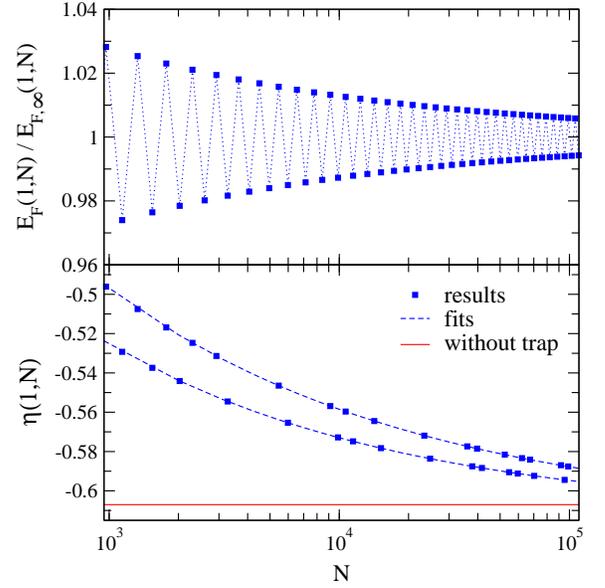}
\end{center}
\caption{(Color online)
Upper panel: Local Fermi energy $E_{\rm F}(1,N)$ 
at the center of an isotropic trap divided by the large-$N$ 
expression $E_{{\rm F},\infty}(1,N)$ as a function of spin-up
particle number $N$. Lower panel: Energy $\eta(1,N)$ for an 
isotropic trap, with fits to our numerical results (see text).
The horizontal
line represents $\eta=-0.607$ for the uniform 
system~\cite{Chevy1}.\label{etavsN}}
\end{figure}

Minimizing $\langle \psi | H | \psi
\rangle$ with respect to $\phi_{0}$,
$\phi_{{\bf m},{\bf h},{\bf p}}$, 
we find the self-consistent equation for $E$ in 
anisotropic traps,
\begin{multline}
E-\varepsilon_{\bf 0} = \\
\sum_{\varepsilon_{{\bf h}} \leqslant \ef} \sum_{{\bf S},{\bf L}}
F({\bf 0},{\bf h},{\bf S}) \, \bigl[ 
M^{-1}(\ef, E + \varepsilon_{{\bf h}}) \bigr]_{{\bf S},{\bf L}} \,
F({\bf 0},{\bf h},{\bf L}) \,,
\label{gain}
\end{multline}
where $E$ is measured from the energy of the Fermi sea,
in weak coupling $E \approx \varepsilon_{\bf 0}$, and
the matrix $M$ is given by
\begin{multline}
M(\ef, E+\varepsilon_{\bf h})_{{\bf S},{\bf L}} 
= \biggl[ \frac{1}{C(\Lambda)}
- D(\alpha, \Delta\widetilde{E}) \biggr] \, \delta_{{\bf S},{\bf L}} \\
+ \sum_{\varepsilon_{\bf p} \leqslant \ef} \sum_{\bf m} 
\frac{F({\bf m},{\bf p},{\bf S}) F({\bf m},{\bf p},{\bf L})}{
E+\varepsilon_{\bf h}-(\varepsilon_{{\bf p}}+\varepsilon_{{\bf m}})} \,.
\label{Mdef}
\end{multline}
Here $\Delta\widetilde{E}= \alpha(S_x+S_y+2)+S_z+1- 
(E+\varepsilon_{\bf h})/\omega$ and 
$D(\alpha, \Delta\widetilde{E})$ is identical to 
the last term of Eq.~(\ref{Mdef})
with unrestricted sum over ${\bf p}$ and ${\bf S}={\bf L}$.
%Similar generalizations as
%Eqs.~(\ref{gain}) and~(\ref{Mdef}) are found for $\phi_0$ and
%$\phi_{{\bf m},{\bf h},{\bf p}}$ compared to the uniform 
%results~\cite{Chevy1,Chevy2}.
For an isotropic trap, $D(1, \Delta\widetilde{E})$
has the simple analytical form
\be
D(1, \Delta\widetilde{E}) = \frac{m \Lambda}{
2(2\pi)^{3/2}} 
+ \biggl( \frac{m\omega}{2\pi} \biggr)^{3/2}
\frac{\sqrt{\pi} \: \Gamma(\Delta\widetilde{E}/2)}{\omega \,
\Gamma((\Delta\widetilde{E}-1)/2)} \,.
\label{Diso}
\ee
The cancellation of the first term in Eq.~(\ref{Diso}) with the
cutoff in the $1/C(\Lambda)$ term in Eq.~(\ref{Mdef}) demonstrates
that $E$ is cutoff independent for large $\Lambda$. We have
verified that this is the case for all studied 
$\alpha$ and use $\Lambda > 10^4 \sqrt{m \omega/2}$.
Moreover, we have found numerically 
that the diagonal matrix elements of $M(\ef, E+\varepsilon_{\bf h})$
depend only on the center-of-mass excitation $\alpha(S_x+S_y)+S_z$.

For large scattering lengths, $1/a_{\rm s} = 0$,
the energy is a universal function
of the aspect ratio and the spin-up particle number $N=N_\uparrow$,
and we generalize
the scaling for the uniform system to anisotropic traps,
\be
E = \eta(\alpha,N) \, E_{\rm F}(\alpha,N) \,,
\label{eta}
\ee
where $E_{\rm F}(\alpha,N) = (6 \pi^2 n_{\uparrow}(0))^{2/3}/(2m)$ is the
local Fermi energy of spin-up particles at the center of the trap.
In the upper panel of Fig.~\ref{etavsN}, we show $E_{\rm F}(1,N)$ for
an isotropic trap divided by the large-$N$ expression 
$E_{{\rm F},\infty}(1,N) = \omega \, (6 N)^{1/3}$. The points are for 
alternating odd-even values of the Fermi level $n_{\rm F}$, which
defines the Fermi energy $\ef = \omega (\alpha n_{\rm F} + (2\alpha+1)/2)$.
The local Fermi energy approaches $E_{{\rm F},\infty}(1,N)$
from above (below) for odd (even) $n_{\rm F}$. With
increasing $\alpha$, this effect decreases and the
envelopes approach the large-$N$ result faster.

\begin{figure}[t]
\begin{center}
\includegraphics[scale=0.45,clip=]{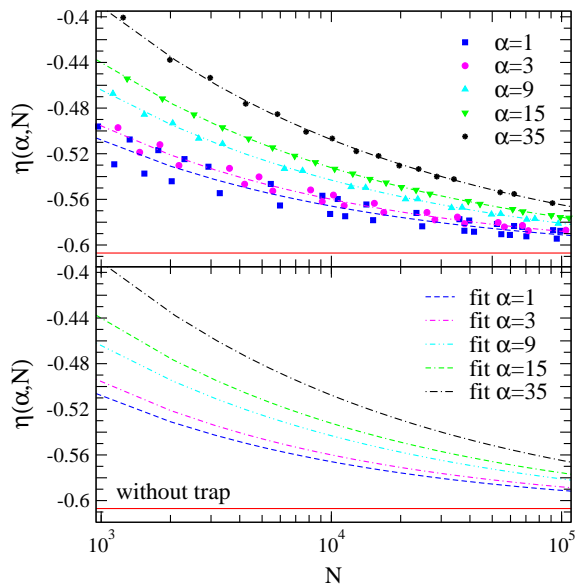}
\end{center}
\caption{(Color online)
Energy $\eta(\alpha,N)$ as a function of 
$N$ for various aspect ratios $\alpha$, compared to 
$\eta=-0.607$ for the uniform system~\cite{Chevy1}
(horizontal lines). The fits are 
discussed in the text and shown separately in the lower panel.
\label{etavsr}}
\end{figure}

{\it Results.}
Using Eq.~(\ref{eta}), we solve Eq.~(\ref{gain}) iteratively for
$\eta(\alpha,N)$, with a numerical precision better than 
$1\%$. To this end, we take the matrix $M(\ef,E+\epsilon_{\bf h})$ to 
be diagonal. This is correct in the large-$N$ limit, and we have 
checked numerically that the off-diagonal matrix elements are
considerably smaller than the diagonal ones for all studied values
of $N$. In the lower panel of Fig.~\ref{etavsN}, we show
$\eta(1,N)$ for an isotropic trap as a function of $N$.
The effects due to finite particle numbers
and confinement of the trap are clearly present: 
The odd-even systematics seen in the
local Fermi energy $E_{\rm F}(1,N)$ is small compared to the decrease
of $\eta(1,N)$ with particle number. Therefore, the decrease
is not due to the change in the local Fermi energy.
For one spin-up fermion,
the exact ground-state energy in a trap~\cite{Busch} is $E(1,1)=
\omega/2 > 0$, thus $\eta(1,1) = (36 \pi)^{-1/3} = 0.207$~\footnote{The 
exact two-body
ground-state energy~\cite{Busch} is reproduced, if we generalize
the first term in Eq.~(\ref{Ansatz}) to include a sum over the
spin-down particle in level ${\bf n}$, $\sum_{\bf n} \phi_{\bf n} 
| {\bf n} \rangle$.}.
With increasing $N$, $\eta(1,N)$
decreases and saturates. Using the Ansatz, 
$\eta(\alpha,N)=a(\alpha) (1+b(\alpha) N^{-c(\alpha)})$, we fit our
numerical results for odd (even) $n_{\rm F}$ separately
and find $a(1) \approx -0.61$ and $c(1) \approx
0.34~(0.32)$. This is in very good agreement with $\eta=-0.607$ for the 
uniform system~\cite{Chevy1} and natural large-$N$ corrections of
$1/E_{{\rm F},\infty}(1,N) \sim N^{-1/3}$. Therefore,
$\sim 10\%$ changes of $\eta$ are natural for $N \sim 10^4$.

In Fig.~\ref{etavsr}, we show the dependence of $\eta(\alpha,N)$
on trap geometry, for various aspect ratios from $\alpha=1$ 
to $\alpha=35$, as a function of the spin-up particle number.
The single-particle energy depends significantly on the aspect
ratio, while the odd-even Fermi level effect decreases with
increasing $N, \alpha$ and is negligible for $\alpha \gtrsim 10$.
For fixed $N$, $\eta(\alpha,N)$ increases with increasing
$\alpha$. In addition, for larger aspect ratios, the dependence
on $N$ is stronger. For each $\alpha$, we fit our combined results 
(including odd and even $n_{\rm F}$) with the power-law
Ansatz and show the fits in Fig.~\ref{etavsr}. We find $a(\alpha)
\approx -0.61(1)$, consistent with the uniform result for all studied
aspect ratios, and $c(\alpha)$ ranges from $c(1) \approx 0.36$ 
to $c(35) \approx 0.31$.

\begin{figure}[t]
\begin{center}
\includegraphics[scale=0.45,clip=]{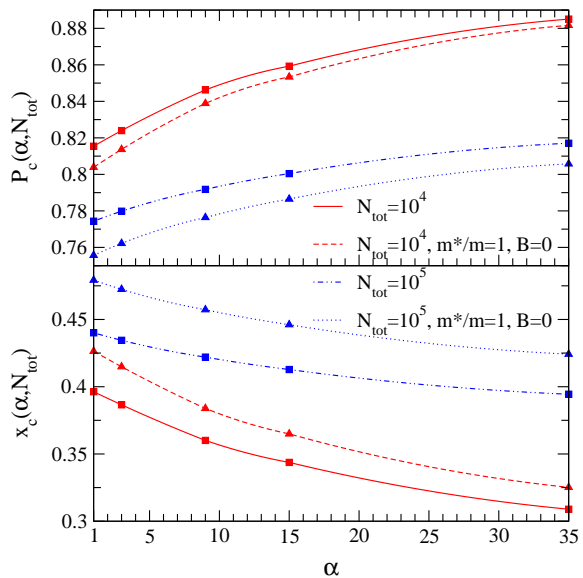}
\end{center}
\caption{(Color online)
Upper panel: Critical polarization $P_c(\alpha,\ntot)$ as a function 
of aspect ratio $\alpha$ for
$\ntot=10^4$ (upper) and $\ntot=10^5$ (lower set of curves).
Lower panel: Critical density ratio $x_c(\alpha,\ntot)$ for $\ntot=10^4$
(lower) and $\ntot=10^5$ (upper set of curves). Results are
shown for two approximations to the quasiparticle spectrum
and interaction.\label{Pcritvsr}}
\end{figure}

{\it Critical polarization.}
We now explore the impact of the calculated finite-size and 
confinement (trap) 
effects on the phase structure.
We consider an unpolarized superfluid
phase and a partially-polarized normal Fermi liquid. Following
Ref.~\cite{Recati}, the free energy is given by
\begin{align}
E_{\rm tot} &= 2 \int_{|{\bf r}| < R_S} \bigl[ \epsilon_{\rm s}(n_{\rm s}({\bf r})) 
+ V({\bf r}) - \mu_{\rm s} \bigr] \, n_{\rm s}({\bf r}) \, d{\bf r} \nonumber \\
&+ \int_{R_{\rm S} < |{\bf r}| < R_{\uparrow}} \bigl[ \epsilon_{\rm n}
(x({\bf r})) \, n_{\uparrow}({\bf r}) 
+ V({\bf r}) \bigl(n_{\downarrow}({\bf r}) 
+ n_{\uparrow}({\bf r})\bigr) \nonumber \\
&\hspace*{21mm}- \mu_{\uparrow} \, n_{\uparrow}({\bf r}) - 
\mu_{\downarrow} \, n_{\downarrow}({\bf r}) \bigr] \, d{\bf r} \,,
\label{energy}
\end{align}
where $x = n_{\downarrow}/n_{\uparrow} \leqslant 1$, 
$R_{\rm S}^2 = \alpha (R_x^2+R_y^2)+R_z^2$ defines 
the boundary of the superfluid phase, and 
the excess spin-up density vanishes at $R_{\uparrow}$. As discussed,
the LDA of Eq.~(\ref{energy}) breaks down for the Rice 
experiment~\cite{Hulet1,Hulet2}. We only use this
here to explore the impact of $\eta(\alpha,N)$ on
the critical polarization. In a full density-functional calculation,
this can also be combined with surface tension~\cite{Mueller}
or gradient terms. For the uniform system at unitarity,
the energy density of the superfluid $\epsilon_{\rm s}$
and of the partially-polarized normal Fermi liquid $\epsilon_{\rm n}$
are given by~\cite{Recati}
\be
\epsilon_{\rm s}(n_{\rm s}) = \xi \, \frac{3}{5} 
\frac{(6 \pi^2 n_{\rm s})^{2/3}}{2m} \text{ and }
\epsilon_{\rm n}(x) = \frac{3}{5} 
\frac{(6 \pi^2 n_\uparrow)^{2/3}}{2m} \epsilon(x) \,, \label{esn}
\ee
with superfluid density $n_{\rm s}$ and universal energy $\xi = 0.42$ of 
the symmetric system~\cite{CR,Pilati}, which is consistent with
$\xi = 0.46 \pm 0.05$ of Ref.~\cite{Hulet1}. Assuming $x \ll 1$, the energy 
of adding spin-down fermions to the normal phase is determined by 
$\eta(\alpha,N)$, 
with corrections due to a spin-down quasiparticle effective mass $m^*$
and due to quasiparticle interactions $B$~\cite{Recati}:
\be
\epsilon(x) =
\biggl[ 1 + \frac{5}{3} \, \eta(\alpha,N) \, x + \frac{m}{m^*} \, x^{5/3}
+ B \, x^2 \biggr] \,.
\label{epsilon}
\ee
We take $\eta(\alpha,N)$ from Fig.~\ref{etavsr}, but for
simplicity consider two cases for the quasiparticle spectrum: $m^*/m = 1$,
$B=0$, as well as the MC values $m^*/m = 1.09$, $B=0.14$~\cite{Pilati}, 
which show these are corrections to the leading effects from 
$\eta$. This however does not include the effects of Fermi 
statistics of the minority particles on $\eta$.

The critical polarization $P_c(\alpha,\ntot)$ is obtained, when the
phase boundary reaches the trap 
center $R_{\rm S} \to 0$. In chemical equilibrium, $\mu_S=(\mu_{\downarrow}
+\mu_{\uparrow})/2$, the ground state of the system is determined by
requiring that the energy functional, Eq.~(\ref{energy}), is stationary
with respect to variations of the densities and of the
phase boundary $R_{\rm S}$, so that the pressure between the two phases 
is equal: $2 n_{\rm s}^2 (\partial \epsilon_{\rm s}/\partial n_{\rm s}) =
n_{\uparrow}^2 (\partial \epsilon_{\rm n}/\partial n_{\uparrow})
+ n_{\uparrow} n_{\downarrow} (\partial \epsilon_{\rm n}/\partial
n_{\downarrow})$. This leads to an equation for the critical density
ratio at the center of the trap, $x_c(\alpha,\ntot)
=x(R_{\rm S}=0,\alpha,\ntot)$~\cite{Recati}:
$\epsilon(x_c) + \frac{3}{5} \, (1-x_c) \, 
\partial_x \epsilon(x_c)
- (2 \xi)^{3/5} \, \epsilon(x_c)^{2/5} = 0$.
Given $\ntot$ and $x_c$,
the spin-up/spin-down densities and particle
numbers are determined from the variation of $E_{\rm tot}$.
%with respect to the densities.

In Fig.~\ref{Pcritvsr}, we show the dependence of the critical 
polarization $P_c(\alpha,\ntot)$ and the critical density ratio
$x_c(\alpha,\ntot)$ as a function of aspect ratio, for total particle
numbers $\ntot=10^4$ and $\ntot=10^5$, where the experimental
differences from the uniform system exists. For $\alpha=1$, 
$\ntot=10^7$ and the MC
$m^*$, $B$ values, we reach the uniform system $P_c=0.74$ and 
$x_c=0.47$. For fixed $\ntot=10^4$ 
and increasing $\alpha$ from
$1$ to $35$, $P_c$ increases from $0.82$ to $0.89$ and $x_c$
decreases from $0.40$ to $0.31$ (for the MC 
$m^*$, $B$ values). For fixed $\alpha=35$, $P_c$ increases with
decreasing $\ntot=10^5, 10^4, 10^3$ from $0.82$, $0.89$, $0.96$
and $x_c$ decreases from $0.39$, $0.31$, $0.18$ (for the MC 
$m^*$, $B$ values; results for $\ntot=10^3$ not shown in 
Fig.~\ref{Pcritvsr}). In addition,
we show in Fig.~\ref{Pcritvsr} the dependence on
the quasiparticle spectrum (through $m^*$) and interaction $B$.
For given $\alpha$ and $\ntot$, $P_c$ is larger and $x_c$ smaller
for the MC $m^*$, $B$ values, compared to $m^*/m=1$, $B=0$,
but as expected, the uncertainty due to $m^*$, $B$ is smaller
than the variation of $P_c$ and $x_c$ with $\alpha$, $\ntot$.
This dependence also becomes weaker with increasing 
$\alpha$ and decreasing $\ntot$.

In summary, for lower particle numbers and more elongated traps,
the energy of the normal polarized phase increases and the 
superfluid extends to larger population imbalances. This
provides a microscopic understanding of the MIT-Rice differences
due to the dependence of the polaron energy on the particle number
and the trap geometry. Finite-size effects are stronger in
highly-elongated systems as the dimensionality of the problem
is continuously reduced with increasong aspect ratio. The
$N+1$-body problem is a natural first step towards general
asymmetries and towards contributions to the total energy
beyond $\eta(\alpha,N)$. In addition, effects from a full 
density-functional calculation need to be studied.

\begin{acknowledgments}
We thank M.\ M.\ Forbes, R.\ J.\ Furnstahl, R.\ Hulet, C.\ J.\
Pethick, T.\ Schaefer
and M.\ Zwierlein for useful discussions. This work was supported in 
part by the NSERC and by the NRC of Canada.
\end{acknowledgments}


\begin{thebibliography}{99}
\bibitem{Zwier1} M.\ W.\ Zwierlein {\it et al.},
%, A.\ Schirotzek, C.\ H.\ Schunck,
%and W.\ Ketterle,
Science {\bf 311}, 492 (2006).

\bibitem{Hulet1} G.\ B.\ Partridge {\it et al.},
%, W.\ Li, R.\ I.\ Kamar, Y.-a.\ Liao, 
%and R.\ G.\ Hulet, 
Science {\bf 311}, 503 (2006).

\bibitem{Zwier2} M.\ W.\ Zwierlein {\it et al.},
%, C.\ H.\ Schunck, A.\ Schirotzek, 
%and W.\ Ketterle, 
Nature {\bf 442}, 54 (2006).

\bibitem{Shin1} Y.\ Shin {\it et al.},
%, M.\ W.\ Zwierlein, C.\ H.\ Schunck, A.\ Schirotzek, 
%and W.\ Ketterle, 
\prl {\bf 97}, 030401 (2006).

\bibitem{Hulet2} G.\ B.\ Partridge {\it et al.},
%, W.\ Li, Y.\ A.\ Liao, R.\ G.\
%Hulet, M.\ Haque, and H.\ T.\ C.\ Stoof, 
\prl {\bf 97}, 190407 (2006).

\bibitem{Schunck} C.\ H.\ Schunck {\it et al.},
%, Y.\ Shin, A.\ Schirotzek, 
%M.\ W.\ Zwierlein, and W.\ Ketterle, 
Science {\bf 316}, 867 (2007).

\bibitem{Shin2} Y.-i.\ Shin {\it et al.},
%, C.\ H.\ Schunck, A.\ Schirotzek, 
%and W.\ Ketterle, 
Nature {\bf 451}, 689 (2008).

\bibitem{NSAC} DOE/NSF NSAC Long Range Plan, {\it The Frontiers of 
Nuclear Science} (2007), p.\ 64-67 and 135-137.

\bibitem{Chevy1} F.\ Chevy, Phys.\ Rev.\ A {\bf 74}, 063628 (2006).

\bibitem{Chevy2} F.\ Chevy, in {\it Ultra-Cold Fermi Gases}, 
%Proceedings of the International School of Physics ``Enrico Fermi''
%Course CLXIV, 
Eds. M.\ Inguscio, W.\ Ketterle, C.\ Salomon, 
p.\ 607 (IOS Press, Amsterdam, 2007); cond-mat/0701350.

%\bibitem{Clogston} A.\ M.\ Clogston, \prl {\bf 9}, 266 (1962).

%\bibitem{Chandra} B.\ S.\ Chandrasekhar, Appl.\ Phys.\ Lett.\ {\bf 1},
%7 (1962).

\bibitem{BF} A.\ Bulgac and M.\ M.\ Forbes, Phys.\ Rev.\ A {\bf 75},
031605(R) (2007).

%\bibitem{Combescot1} R.\ Combescot, A.\ Recati, C.\ Lobo, and
%F.\ Chevy, Phys.\ Rev.\ Lett.\ {\bf 98}, 180402 (2007).

\bibitem{Lobo} C.\ Lobo {\it et al.},
%, A.\ Recati, S.\ Giorgini, and S.\ Stringari,
Phys.\ Rev.\ Lett.\ {\bf 97}, 200403 (2006).

\bibitem{CR} J.\ Carlson and
S.\ Reddy, Phys.\ Rev.\ Lett.\ {\bf 95}, 
060401 (2005).

\bibitem{PS} N.\ Prokof'ev and B.\ Svistunov, Phys.\ Rev.\ B {\bf 77},
020408(R) (2008); {\it ibid.} {\bf 77}, 125101 (2008).

\bibitem{Pilati} S.\ Pilati and S.\ Giorgini, Phys.\ Rev.\ Lett.\ {\bf 100}, 
030401 (2008).

\bibitem{Combescot2} R.\ Combescot and S.\ Giraud, Phys.\ Rev.\ Lett.\ 
{\bf 101}, 050404 (2008).

\bibitem{Busch} T. Busch {\it et al.},
%, B.-G.\ Englert, K.\ Rzazewski, and
%M.\ Wilkens, 
Found. Phys. {\bf 28}, 549 (1998).

\bibitem{Mueller} T.\ N.\ De Silva and E.\ J.\ Mueller, \prl {\bf 97},
070402 (2006).

\bibitem{Recati} A.\ Recati, C.\ Lobo, and S.\ Stringari, Phys.\ Rev.\
A {\bf 78}, 023633 (2008).
\end{thebibliography}
\end{document}